\documentclass{article}
\usepackage{arxiv}
\usepackage[utf8]{inputenc} 
\usepackage[T1]{fontenc}    
\usepackage{hyperref}       
\usepackage{url}            
\usepackage{booktabs}       
\usepackage{amsfonts}       
\usepackage{nicefrac}       
\usepackage{microtype}      
\usepackage{lipsum}
\usepackage{graphicx}
\usepackage{booktabs}       
\usepackage{amsfonts}       
\usepackage{nicefrac}       
\usepackage{microtype}      
\usepackage{lipsum}
\usepackage{graphicx}
\usepackage{amsmath,amssymb}
\usepackage{dcolumn}
\usepackage{bm}
\usepackage{braket}
\usepackage[dvipsnames]{xcolor}
\usepackage{xcolor,colortbl}
\usepackage{makecell,booktabs}
\usepackage{graphicx}
\usepackage{dcolumn}
\usepackage{listings}
\usepackage{bm}
\usepackage{braket}
\usepackage[dvipsnames]{xcolor}
\usepackage{xcolor,colortbl}
\usepackage{braket}
\usepackage{amsmath,amssymb,amsfonts}
\usepackage{algorithmic}
\usepackage{textcomp}
\usepackage{url}
\usepackage{bm}
\usepackage{times}
\usepackage{epsfig}
\usepackage{graphicx}
\usepackage{amsmath}
\usepackage{multirow}
\usepackage{amssymb}
\usepackage{subfig}
\usepackage{graphicx}
\usepackage{float}
\usepackage{booktabs}
\usepackage{textcomp}
\usepackage{adjustbox}\usepackage{mathtools}
\usepackage{array}
\usepackage{longtable}
\usepackage{makecell,booktabs}
\usepackage{multirow}
\usepackage{stackengine}
\usepackage{algorithm}
\usepackage{graphicx}
\usepackage{listings}
\usepackage{color}
\usepackage{color,array}
\usepackage{cite}
\usepackage{amsmath,amssymb,amsfonts}
\usepackage{algorithmic}
\usepackage{textcomp}
\usepackage{url}
\usepackage{bm}
\usepackage{times}
\usepackage{epsfig}
\usepackage{graphicx}
\usepackage{amsmath}
\usepackage{multirow}
\usepackage{amssymb}
\usepackage{subfig}
\usepackage{graphicx}
\usepackage{float}
\usepackage{booktabs}
\usepackage{textcomp}
\usepackage{adjustbox}\usepackage{mathtools}
\usepackage{array}
\usepackage{makecell,booktabs}
\usepackage{multirow}
\usepackage{stackengine}
\usepackage{algorithm}
\usepackage{graphicx}
\usepackage[dvipsnames]{xcolor}
\usepackage{xcolor,colortbl}
\newtheorem{definition}{Definition}
\graphicspath{ {./images/} }
\usepackage{listings}
\usepackage[title]{appendix}
\usepackage{rotating}
\definecolor{amber}{rgb}{1.0, 0.75, 0.0}
\definecolor{orange}{rgb}{1.0, 0.49, 0.0}
\definecolor{codegreen}{rgb}{0,0.6,0}
\definecolor{codegray}{rgb}{0.5,0.5,0.5}
\definecolor{codepurple}{rgb}{0.58,0,0.82}
\definecolor{backcolour}{rgb}{0.95,0.95,0.92}

\lstdefinestyle{mystyle}{
    backgroundcolor=\color{backcolour},   
    commentstyle=\color{codegreen},
    keywordstyle=\color{magenta},
    numberstyle=\tiny\color{codegray},
    stringstyle=\color{codepurple},
    basicstyle=\ttfamily\footnotesize,
    breakatwhitespace=false,         
    breaklines=true,                 
    captionpos=b,                    
    keepspaces=true,                 
    numbers=left,                    
    numbersep=5pt,                  
    showspaces=false,                
    showstringspaces=false,
    showtabs=false,                  
    tabsize=2
}
\title{Quan\textit{Ants} Machine: A Quantum Algorithm for Biomarker Discovery}
\author{
 Phuong-Nam Nguyen \\
  Tampa, FL 33620 \\
  \texttt{namphuongnguyen510@gmail.com} \\
}
\begin{document}
\maketitle

\begin{abstract}
    The discovery of biomarker sets for a targeted pathway is a challenging problem in biomedical medicine, which is computationally prohibited on classical algorithms due to the massive search space. Here, I present a quantum algorithm named Quant\textit{Ants} Machine to address the task. The proposed algorithm is a quantum analog of the classical Ant Colony Optimization (ACO). We create the mixture of multi-domain from genetic networks by representation theory, enabling the search of biomarkers from the multi-modality of the human genome. Although the proposed model can be generalized, we investigate the \textit{RAS}-mutational activation in this work. To the end, Quant\textit{Ants} Machine discovers rarely-known biomarkers in clinical-associated domain for \textit{RAS}-activation pathway, including \textit{COL5A1}, \textit{COL5A2}, \textit{CCT5}, \textit{MTSS1} and \textit{NCAPD2}. Besides, the model also suggests several therapeutic-targets such as \textit{JUP}, \textit{CD9}, \textit{CD34} and \textit{CD74}.
\end{abstract}

\section{Introduction}\label{sec:introduction}
Cancer is a causality process. Specifically, given a patient $\rho$ with the genetic dataset $\mathbb{G} = \{G_1,\dots, G_n \}$, the mutation in a single biomarker could be caused by the activation of some other biomarkers, which can be presented as a causality model. In this paper, the \textbf{biomarker} is defined as: \textit{"A substance in the blood, bodily fluids, or tissues indicating a typical or atypical physiological process, state, or state ailment.\cite{nci-biomarker}"}. We present the event $E_i$ that a gene $G_i$ witnesses a mutational event. Of note, mutations happen constantly\cite{lewontin1985population}; hence, biomarkers attributed to a single \textbf{target} could vary. However, the underlying relationship among genes (still obscure up to this point) could leave the trait of evidence, with can be presented statistically. We define \textbf{targets} as biomarkers with empirical evidence that certainly leads to carcinogenesis. We will use the \textit{RAS}-family (a group of \textit{oncogenes}) for demonstrating the proposed algorithm. 

We denote target as $\bm{Y}$, in which $\mathbb{P}(\bm{Y} = 0) = p$ and $\mathbb{P}(\bm{Y} = 1) = 1-p$; where $\{\bm{Y} = 1\}$ is the event that the target got mutational activation; otherwise the biomarker-target stay wild-type. We use the term biomarker-target because a target can be a biomarker of other targets.

We define the \textbf{biomarker set} that activates the mutation on the target $\bm{Y}$ as $\mathcal{G} = \{G_i,G_j, \dots,G_k\}$; $i,j,k \in \text{Index}$ set of the mother set $\mathbb{G}$. The complexity of finding this subset \textbf{structure} is challenging. Specifically, there are $k \choose n$ number of such subsets. Here, we highlight that each person will have characterized patterns for mutational alterations, which induce idiosyncratic biomarker sets. For all possibilities, we have the summation combination of
\begin{equation}
    \sum_{i=0}^{N}{N\choose k} = 2^N
\end{equation}
possibilities. Human genomes have approximately $N=20,000$ to $25,000$ genes which induce a massive search space of $3.2019 \times 10^{6577}$ candidate biomarker sets for an \textbf{biomarker identification algorithm}, exponentially scaled with the number of input genes. Here, we would like to define the idiosyncratic-biomarker set of a given cohort as the \textbf{biomarker-verse}, as the number of possibilities is much larger than the number of atoms in one universe\cite{nielsen2002quantum} ($2^{300}$-$2^{500}$).

This article introduces Quant\textit{Ants} machine, a quantum algorithm for the discovery of a biomarker-verse associated with a given dataset $\mathcal{D} = \{\bm{X},\bm{Y} \}$, in which $\bm{X}$ is the \textbf{feature set}; or \textbf{genetic representations} like DNA Methylation, RNA/miRNA expression, protein expressions. In other words, the outcome of our model is data-driven, which can be implemented in the next generation of quantum computers. Here, we will only introduce the quantum algorithm from the fundamentals of mathematical oncology. We organize the article as follows: \textbf{Section}~\ref{sec:method} introduces the algorithm, Quant\textit{Ants} machine for the identification of data-driven biomarker-verse. \textbf{Section}~\ref{sec:result} reports the numerical results using three case studies associated with the running exemplary of the discussed targets. We will give several future directions to this work in \textbf{Section}~\ref{sec:conclusion} with clinical significance discussed in \textbf{Section}~\ref{sec:clinical_significance}.

\section{Quant\textit{Ants} Machine}\label{sec:method}
\begin{figure}[t]
    \centering
    \includegraphics[width = \textwidth]{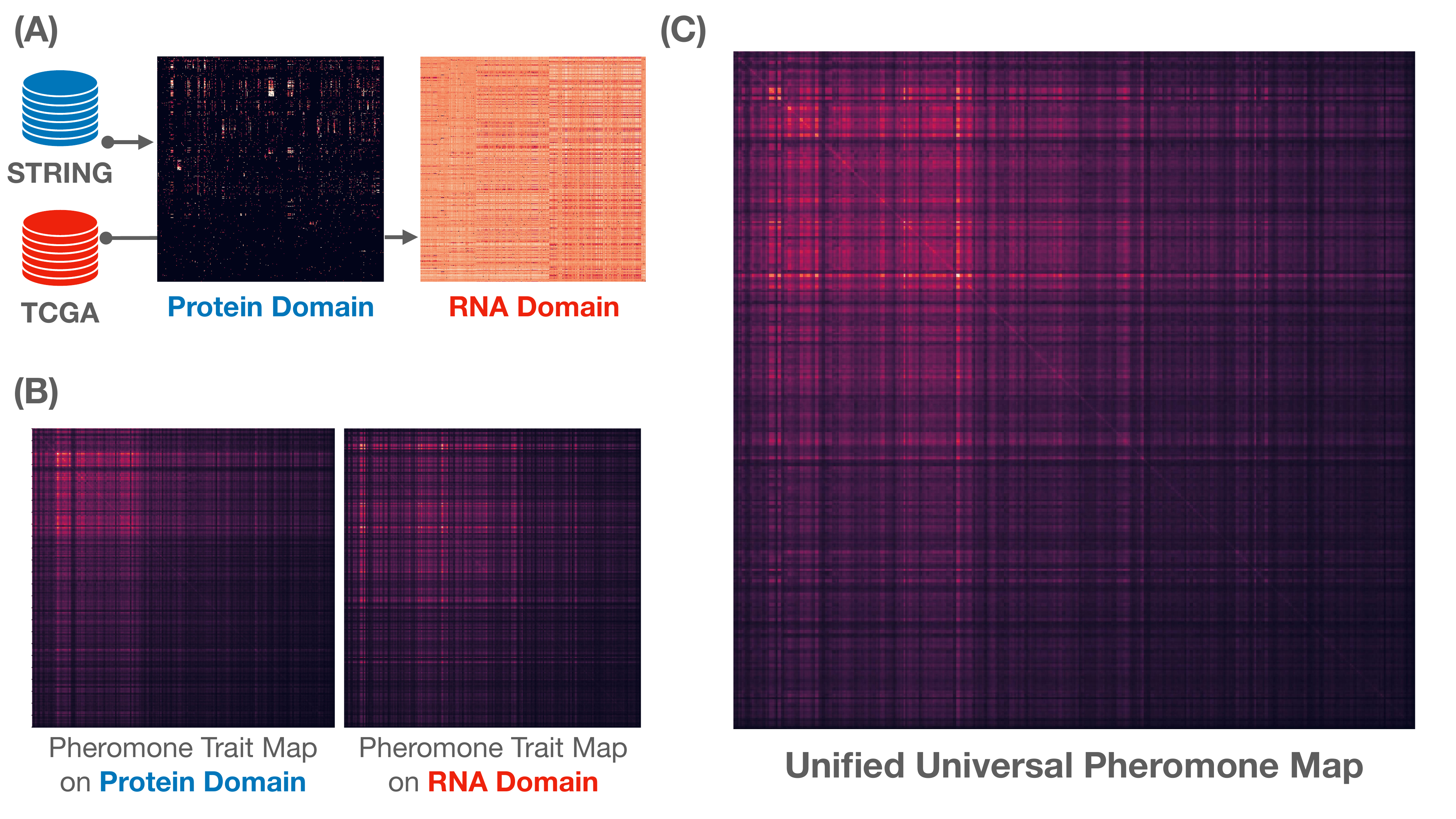}
    \caption{(\textbf{A}) Protein interaction networks from STRING database\cite{szklarczyk2019string} (\textbf{Protein domain}) and RNA statistical graph (\textbf{RNA domain}) computed based on mRMR criteria (\textbf{SuppMat}-~\ref{supp:mrmr}). (\textbf{B}) Pheromone trait maps left by quantum-ant agents after an optimized walk-on among the dual regimes. (\textbf{C}) Unified Universal Pheromone Map (\textbf{UPM}) learned by the Quant\textit{Ants} for Clinical Oncology using four ant agents.}
    \label{fig:maps}
\end{figure}

Given $\bm{A}_b$ and $\bm{A}_s$ are adjacency associated with two domain graphs $\mathbb{B}_0$ and $\mathbb{B}_1$ induced from input dataset $\mathcal{D}$. We will construct dual representations for the two given matrices in \textbf{Figure}~\ref{fig:maps}(A). 
\begin{enumerate}
    \item $\mathcal{G}_{1}$ as a protein interaction network retrieved from STRING\cite{szklarczyk2019string} database (\textbf{Figure}~\ref{fig:maps}). 
    \item $\mathcal{G}_{2}$ as an interaction network computed from RNA expression by mRMR criteria (\textbf{SuppMat}~\ref{supp:mrmr}).
\end{enumerate}

\begin{definition}
    A representation of a group $\mathcal{G}$ is a homomorphism $\Phi: \mathcal{G} \rightarrow GL(V)$  with $GL(V):= \{f \in End(V)| f\textbf{ is invetible}\}$. We have $End(V)$ as the endomorphism ring of (finite-dimensional) vector space $V$, or a set of linear mapping from $V$ to $V$.
\end{definition}
\begin{definition}\label{def:dual_reps}
    We defined the mapping with direct sum (from Euler's formula)
    \begin{equation}\label{equa:euler}
        [\Phi^{(0)} \oplus \Phi^{(1)}] = \begin{bmatrix}
        \exp(\frac{2\pi i m}{n}) & 0 \\
        0 & \exp(-\frac{2\pi i m}{n})
        \end{bmatrix}
    \end{equation}
    as \textbf{Euler Representations}\cite{feit1982representation}. The dual-generators for sub-representations is
    \begin{equation}
    \begin{split}
        \Phi^{(0)}_{[m]} &= \exp\bigg( {2\pi i}\frac{m}{n} \bigg)\\
        \Phi^{(1)}_{[m]} &= \exp\bigg({-2\pi i}\frac{m}{n} \bigg).
        \end{split}
    \end{equation}
\end{definition}

We encode the domain information, i.e., $\bm{A}_b$ and $\bm{A}_s$, by altering Euler Representations with spectral decomposition theorem. We have
\begin{equation}
    \begin{split}
    \bm{A}_b &= \bm{S}_b^\dagger \Lambda_b \bm{S}_b\\
    \bm{A}_s &= \bm{S}_s^\dagger \Lambda_s \bm{S}_s
    \end{split}
\end{equation}
where $\Lambda_b = \text{diag} (\lambda^{(b)}_1, \dots, \lambda^{(b)}_N)$ and $\Lambda_s = \text{diag} (\lambda^{(s)}_1, \dots, \lambda^{(s)}_N)$ are diagonal matrix of eigenvalues for $\bm{A}_b$ and $\bm{A}_s$, respectively. Noted that both $\Lambda_b$ and $\Lambda_s$ are complex-valued matrix as $\bm{A}_b$ and $\bm{A}_s$ is asymmetric matrices; i.e., $\bm{A}_b \neq \bm{A}_b^{\dagger}$ and $\bm{A}_s \neq \bm{A}_s^{\dagger}$. 

We define the mapping
\begin{equation}\label{equa:dualmap}
\begin{split}
    \Phi^{(0)}_{[\Delta t]}(\lambda^{(b)}_k) &= \exp\bigg(i\Delta t \lambda^{(b)}_k \bigg)\\
    \Phi^{(1)}_{[\Delta t]}(\lambda^{(s)}_k) &= \exp\bigg(i\Delta t \lambda^{(s)}_k \bigg)
    \end{split}
\end{equation}
generating the dual representations:
\begin{equation}\label{equa:lambda_phi}
    \sum_{k=1}^{n} \Phi^{(0)}_{[\Delta t]}(\lambda^{(b)}_k) \otimes \Phi^{(1)}_{[\Delta t]}(\lambda^{(s)}_k) =i\Delta t \times \text{diag} (\lambda^{(b)}_0 + \lambda^{(s)}_0, \dots, \lambda^{(b)}_N + \lambda^{(s)}_N) := \Lambda(\Phi[\Delta t])
\end{equation}

We now have the initial transition maps $\Lambda(\Phi[\Delta t])$ containing information for both protein and RNA domains; i.e., $\bm{A}_b$ and $\bm{A}_s$ are presented into $\Lambda(\Phi)[\Delta t]$ through their eigenvalue set.

\begin{definition}\label{def:quan_propa}
We define \textbf{quantum propagation} from one state to other-state in a short time step $\Delta t$ as
\begin{equation}~\label{equa:Ut} 
    \begin{split}
        \ket{\psi_t} &= \bm{U}_r(\Delta t) \ket{\psi({t_j})}\\
        \bm{U}_{r}(\Delta t) &= \bm{S}_{r}^\dagger \Lambda[\Phi (\Delta t)] \bm{S}_{r},
    \end{split}
\end{equation}
with $r$ is either associated with $\mathbb{B}_0$ or $\mathbb{B}_1$. We define a simple \textbf{propagator} (transformation from complex-valued to real-valued) as the $|\bm{U}_{r}(\Delta t)|:= \{ u_{ij} = \sqrt{a^2+b^2}| u_{ij} = a+bi; (a,b) \in \mathbb{R}\times \mathbb{R}\}.$
\end{definition}

\subsection{Algorithms}

We introduce Quant\textit{Ants} training procedure in \textbf{Algorithm}~\ref{algo:quantants}. Our algorithm has emergent properties: with some choice for $\Delta t$, the quantum state after propagation processes can converge to the initial state within the loop. The quantum agents will finish discovering the dual domains at some point. There is some crucial observation surrounding \textbf{Algorithm}~\ref{algo:quantants}:
\begin{itemize}    
    \item The proposed quantum propagation in (\textbf{Definition}~\ref{def:quan_propa}) deviates the agent into a dual-agent system walking on two graphs $\bm{A}_b$ and $\bm{A}_s$, represented by the corresponding $\ket{\psi_{\Delta t}^{(b)}}$ and $\ket{\psi_{\Delta t}^{(s)}}$.
        
    \item Thus, each element of the set $\mathcal{A}_b$ and $\mathcal{A}_s$ are consider the \textit{\textbf{pheromone trait maps}} during the quantum walking process (unobservable). In other words, we cannot tell the exact adventure of the ant, but \textit{the pheromone trait maps can hint at which genes the ant visited with some intensity (probability). }
    
    \item By all means, this is a \textbf{\textit{quantum phenomenon}}: we cannot tell the exact location of a sub-particle (the ant) but instead the probability of finding them in some interval of locations. The ant system is walking dependently on the dual regimes, as the domain information is tied (entangled) by representation \textbf{Equation}~\ref{equa:lambda_phi}.
    
\end{itemize}

\begin{algorithm}[t]
  \caption{Quant\textit{Ants}}
  \label{algo:quantants}
  \textbf{Input}: $\bm{A}_b$ and $\bm{A}_s$ are adjacency maps for the protein interaction network $\mathbb{B}_0$ and the RNA domain $\mathbb{B}_1$$.
  \Delta t$ is propagating time.
  \begin{enumerate}
    \item [] \textbf{create}: $\mathcal{A}_b = [\mbox{ }]$, $\mathcal{A}_s = [\mbox{ }]$
    \item[]  \textbf{initiate} $\ket{\psi_0} = \ket{0}_{N} = (1,0,\dots,0)^\intercal$
    \item [] \textbf{compute:} 
    \begin{enumerate}
        \item []  $|\bm{U}_{r}(\Delta t)| (r\in {b,s}$ as in \textbf{Equation}~\ref{equa:Ut})
        \item []  $\ket{\psi_{\Delta t}^{(b)}} = \bm{U}_b(\Delta t)\ket{\psi_0}$
        \item []  $\ket{\psi_{\Delta t}^{(s)}} = \bm{U}_s(\Delta t)\ket{\psi_0}$
    \end{enumerate}    
    \item [] \textbf{for} $t$ in \{1, iters\}:
    \begin{enumerate}
        \item [] \textbf{compute}:
        \begin{enumerate}
            \item [] $\bm{S}_b$, ${\Lambda_b} \leftarrow$ Spectral Decomposition($\bm{U}_b(\Delta t)$)
            \item [] $\bm{S}_s$, ${\Lambda_s} \leftarrow$ Spectral Decomposition($\bm{U}_s(\Delta t)$)
            \item [] ($\bm{S}_r, r\in\{b,s\}$ is eigenvector matrix, $\Lambda_r$ is diagonal matrix of eigenvalues)
        \end{enumerate}
        \item [] \textbf{update}:
        \begin{enumerate}
            \item [] $\Lambda(\Phi[\Delta t])  \leftarrow \sum_{k=1}^{n} \Phi^{(0)}_{[\Delta t]}(\lambda^{(b)}_k) \otimes \Phi^{(1)}_{[\Delta t]}(\lambda^{(s)}_k)$; ($\lambda^{(b)}_k \in \Lambda_b$, $\lambda^{(s)}_k \in \Lambda_s$).
            \item [] $\bm{U}_{r}(\Delta t) \leftarrow \bm{S}^\dagger_r \Lambda(\Phi[\Delta t]) \bm{S}_r, r\in \{b,s\}$
            \item []  $\ket{\psi_{\Delta t}^{(r)}} \leftarrow \bm{U}_r(\Delta t)\ket{\psi_0}, r\in \{b,s\}$
        \end{enumerate}
        \item [] $\mathcal{A}_b$.append($|\bm{U}_{b}(\Delta t)|$)
        \item [] $\mathcal{A}_s$.append($|\bm{U}_{s}(\Delta t)|$)
        \item [] \textbf{end if:} $\ket{\psi_{\Delta t}^{(r)}} = \ket{\psi_0}, r\in \{b,s\}$
    \end{enumerate}
    \item [] \textbf{return}: learned maps $\mathcal{A}_b$ and $\mathcal{A}_s$.
  \end{enumerate}
\end{algorithm}

\begin{algorithm}[t]
    \caption{Multi-Quant\textit{Ants}}
    \label{algo:multi-quantants}
    \textbf{Input}: A k-ant system is initiated by \{ant($i$)\} with each ant parameterized by $\Delta t \in \{ \pi / 11, 2\pi /23, 3\pi/31,4\pi /41, \dots, k\pi/p \}$, with $p$ is the smaller prime number between $[\bar{k}0, \bar{k}9]$.  
    \begin{enumerate}
        \item [] \textbf{create}: $\mathcal{U}_{b} = [\mbox{ }]$ and $\mathcal{U}_{s} = [\mbox{ }]$
        \item [] \textbf{for} $i \in [1, \dots, k]$:
        \begin{enumerate}
            \item [] $\mathcal{A}_b(k)$, $\mathcal{A}_s(k)$ = QuantAnts($k\pi/p$)
            \item [] $\mathcal{U}_{b}$.append($\mathcal{A}_b(k)$)
            \item [] $\mathcal{U}_{s}$.append($\mathcal{A}_b(s)$)
        \end{enumerate}
    \item [] \textbf{return}: universal maps $\mathcal{U}_{b}$ and $\mathcal{U}_{s}$
    \end{enumerate}
\end{algorithm}

The proposed algorithm can then be extended to a multi-agent system, i.e., starting with different ant-agent $\alpha(\Delta t)$ parameterized by troubled times. As discussed, each initial setting will result in completely different pheromone maps; thus, new connections from gene-to-gene can be found when multi-Quant\textit{Ants} searching for biomarkers. We define the algorithm output as \textbf{universal pheromone maps} (\textbf{UPM}) $\mathcal{U}_{b}$ and $\mathcal{U}_{s}$, with
\begin{equation}
    \mathcal{U}_r = \text{aggregation}(\mathcal{A}_r(k)), r\in \{b,s \}.
\end{equation}
Within the scope of this work, we consider simple mean-pooling over the pheromone maps given as 
\begin{equation}\label{equa:umap_mean}
    \mathcal{U}_r = \frac{1}{n_{k}}\sum_{k=1}^{n_{k}}([\mathcal{A}_r(k)]_{[N, N]}), r\in \{b,s \};
\end{equation}
with $n_{k}$ is the number of discovered pheromone maps by the quantum $k$-ant system. We use the SoftMax function to normalize the simulated states at each iteration so that presenting saturated distributions. 

We specifically choose the basis set $\Delta t \in \{ \pi / 11, 2\pi /23, 3\pi/31,4\pi /41, \dots, k\pi/p \}$ generated by prime numbers: the nominator is scalar $k\pi$, while the denominator is the least prime number between $[\bar{k}0, \bar{k}9]$. This selection makes the exponential terms in our dual-generators (\textbf{Equation}~\ref{equa:dualmap}) become
\begin{equation}
    i\Delta t \lambda = i \frac{k \pi}{p}\lambda =2\pi i \frac{k\lambda}{2p}
\end{equation}
The selection of $\Delta t$ from the prime basis provides good ansatzes for (guessed) for \textbf{domain propagation} that makes asymmetric domain to a more symmetric domain; i.e., $\frac{k\lambda}{2p} \rightarrow \frac{m}{n}$ with $\text{gcd}(m,n) = 1$ making asymmetric domain (\textbf{Figure}~\ref{fig:maps}) to the symmetric domain (\textbf{Definition}~\ref{equa:euler}). 
\begin{figure}[t]
    \centering
    \includegraphics[width = \textwidth]{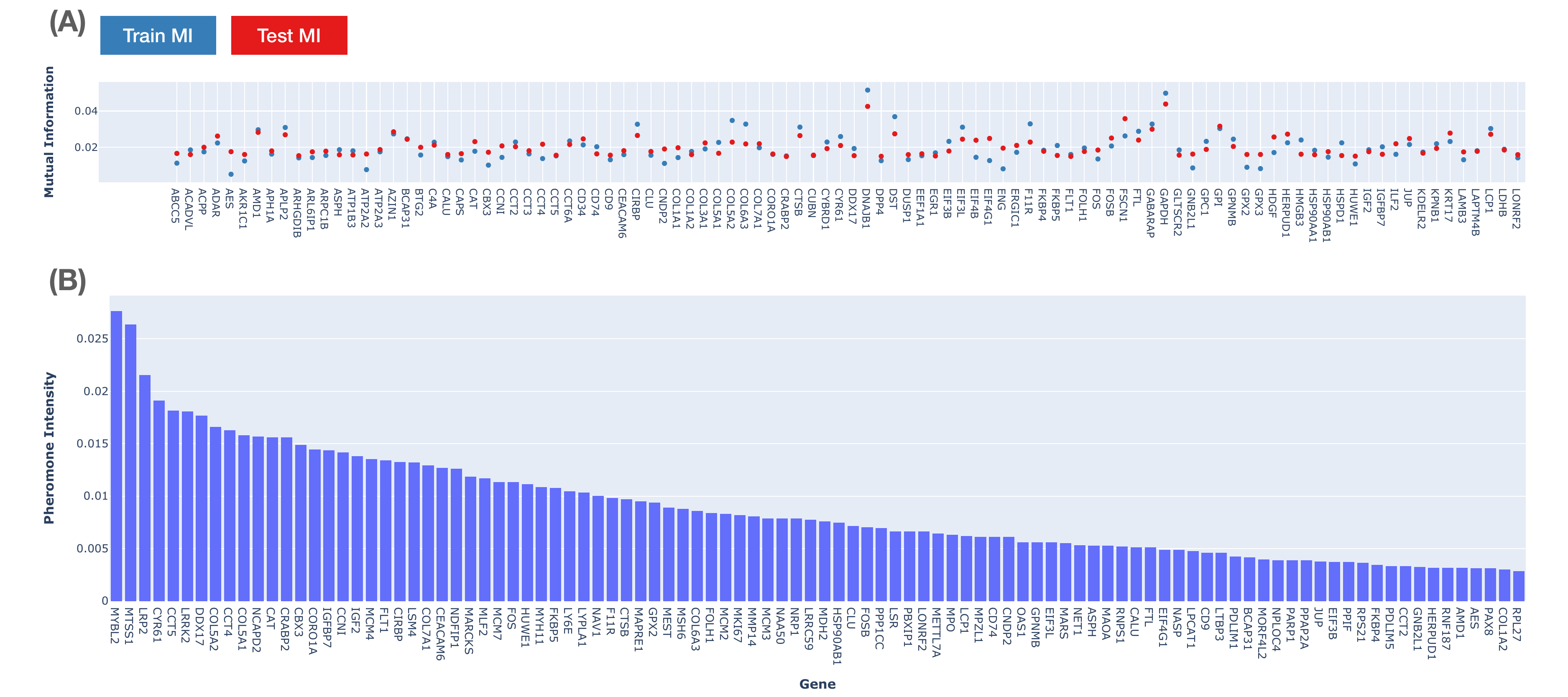}
    \caption{\textbf{(A)} Train/Test Relevancy of Biomarkers towards Mutational Activation of \textit{RAS}-family targets. \textbf{(B)} Biomarkers Score (\textbf{G-Score}) based on Pheromone Intensity by The Quant\textit{Ants} Machine.}
    \label{fig:result_1}
\end{figure}
\begin{table}[t]
    \centering
    \caption{\textbf{Discovered Biomarkers of the \textit{RAS}-family proposed by QuantAnts Machines based on \textbf{G-Score} (pheromone intensity) and relevancy score.} The biomarker with superscript $^*$ is rarely-known in clinical-associated literature (less than $100$ associated publications), addressed through literature mining (\textbf{SuppFig}~\ref{fig:cli_sig}).}
    \scalebox{0.9}{
    \begin{tabular}{|c|c|}
    \toprule
   \textbf{Discovered Biomarkers (Post-Selection)} & \textbf{Catergory}\\
    \midrule
    \textit{COL5A2}$^*$, \textit{COL5A1}$^*$, \textit{COL7A1}, \textit{COL6A3}, & Collagen chains\\
    \textit{COL1A2}, \textit{COL1A1}, \textit{COL3A2}. & \\
    \midrule
    \textit{CD9}, \textit{CD34}, \textit{CD74}. & Immuno-therapeutic antigens \\
    \midrule
    \textit{MYBL2}, \textit{MTSS1}$^*$, \textit{LRP2}, \textit{EGR1}, \textit{GABARAP}, & Other notable cancer-related genes\\
    \textit{HSP90AA1}, \textit{HSP90AB1}, \textit{JUP}, \textit{IGF2}, \textit{MCM3}, &\\
    \textit{MCM4}, \textit{MMP14}, \textit{MCM2}, \textit{MCM4}, \textit{MSH6}, \textit{CCT5}$^*$, \textit{NCAPD2}$^*$ &\\
    \bottomrule
    \end{tabular}}
    \label{tab:res_1} 
\end{table}

\section{Results}\label{sec:result}
\subsection{Case-study}\label{sec:case_study}
\subsubsection{\textit{RAS}-activation Pathway}
The Ras/Raf/MAPK pathway is a signal transduction pathway that is extensively studied in the field of cell biology\cite{knijnenburg2018genomic}. Its main role is to convey signals from outside the cell to the cell nucleus, where specific genes are activated to promote cell growth, division, and differentiation. Additionally, this pathway is crucial in regulating the cell cycle, promoting wound healing and tissue repair, and enabling integrin signaling and cell migration\cite{stacey2003cyclin,boonstra1995epidermal,longhurst1998integrin}. Notably, it also can induce angiogenesis, which involves the formation of new blood vessels, by regulating the expression of genes involved in this process. Overall, the Ras/Raf/MAPK pathway controls various cellular functions that are significant in the development of tumors. In this work, we concern three biomarkers for the \textit{RAS}-family, including \textit{NRAS}, \textit{HRAS}, and \textit{KRAS}. We define the mutation activation of the \textit{RAS}-family as when at least one of the three \textit{RAS}-associated targets got mutations. 

\subsubsection{Datasets}
We use The Cancer Genome Atlas (TCGA\cite{weinstein2013cancer}) for RNA expression data and the STRING\cite{szklarczyk2019string} database for interaction network. The dataset includes $9,136$ samples, with $402$ input genes. We pre-select the top gene with high variation in its expressions. Although the number of evaluated genes is relatively smaller than the whole genome, the search space induced by this evaluated set is huge, equals $2^{402} \approx 1.03 \times 10^{121}$. However, the proposed algorithm is a generalized model that can be applied to any input gene set. Hence, the scalability and robustness of larger databases are worth investigating in further research. The generation of the mRMR-based graph is given in \textbf{Appendix}~\ref{supp:mrmr}.

\subsubsection{Environmental Settings}
Our proposed algorithm is a cost-efficient computational approach. All experiments used Python $3.7.0$, numpy $1.21.5$, sci-kit-learn $1.0.2$, and PyTorch $1.11$ on an Intel i9 processor (2.3 GHz, eight cores), 16GB DDR4 memory and GeForce GTX $1060$ Mobile GPU with 6GB memory. The full technical report is in \textbf{Appendix}~\ref{supp:tech_rep}.

\subsection{Discovered Biomarkers}\label{sec:discovery}
We summarize the biomarker-verses found for targeting the \textit{RAS}-family in \textbf{Figure}~\ref{fig:result_1}. We performed 10-Fold cross-validation to evaluate the relevancy of sampled biomarker sets toward a given target (\textbf{Figure}~\ref{fig:result_1}(A)). The ratio of train/test sets is $80\%/20\%$ with no overlapped patients. The database used in this numerical result is given in \textbf{SuppMat}-S1. The train MI and test MI is positively correlated with $R$-squared of $0.64$ in \textit{RAS}-associated targets (\textbf{SuppFig}~\ref{fig:mrmr_graph}). We associate the pheromone intensity with biomarker score, which is discussed in \textbf{Appendix}~\ref{supp:discussion} and summarized in \textbf{Figure}~\ref{fig:result_1}(B). The top three biomarkers of the \textit{RAS}-family targets proposed by Quant\textit{Ants} Machines is \textit{MYBL2}, \textit{MTSS1} and \textit{LRP2}. We post-select the $27$ markers by two criteria: (1) high phenomenon intensity and (2) high relevancy score to the target. As a result, our study suggests $27$ notable biomarkers grouped in three categories, which is summarized in \textbf{Table}~\ref{tab:res_1}.

\subsection{Clinical Significance}\label{sec:clinical_significance}
\subsubsection{Validation through Literature Mining}
One of the main challenges in the biomarker discovery task is validating the clinical significance of the found biomarkers. Here, we perform a comprehensive literature mining of the discovered biomarker verse. Specifically, we analyze the abstract of $16,661$ publications from \textit{PubMed.gov} regarding $27$ notable markers proposed from our algorithm from 2018 to 2022 (the collections is up to \textit{12/06/2022}). The keywords for such protocol are given as "\textit{Found Biomarkers}" and "\textit{RAS}." The literature mining offers a comprehensive view of the development of biomarker discovery tasks. We highlight five newly discovered drivers for \textit{RAS}-family that suggested by Quan\textit{Ants} Machine, including \textit{MTSS1}, \textit{COL5A1}, \textit{COL5A2}, \textit{CCT5} and \textit{NCAPD2} with less than $100$ associated publications. We summarize the newly discovered biomarkers' biological meanings and related pathways in \textbf{Table}~\ref{tab:gene-pathways}.

\subsubsection{Translation of Algorithm to Potential Therapeutic Targets}
\begin{figure}
    \centering
    \includegraphics[width = \textwidth]{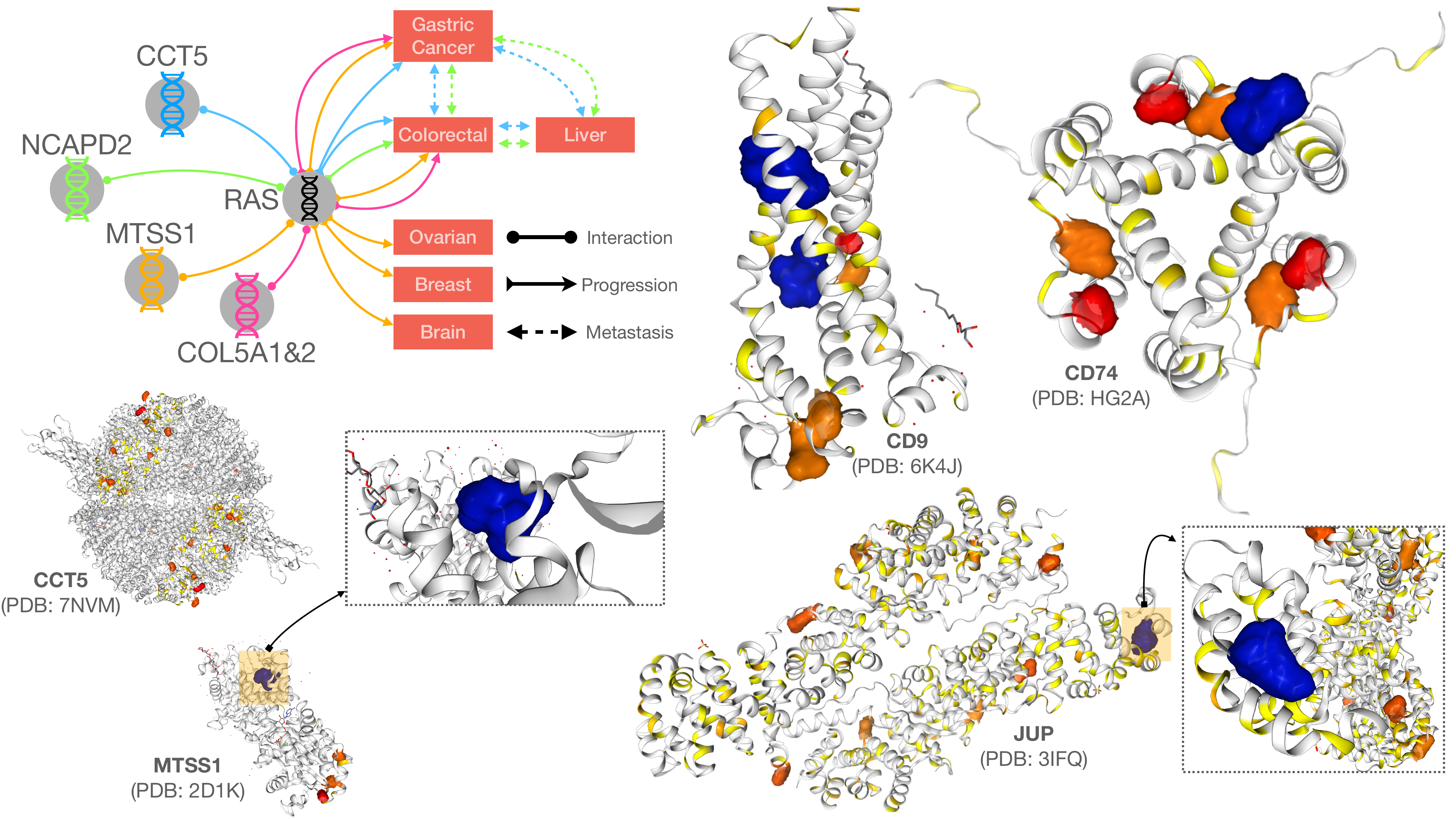}
    \caption{\textbf{Illustrations for Therapeutic Targets discovered from The Biomarker-verse}: (\textbf{Group 1}) \textit{CCT5}, \textit{NCAPD2}, \textit{MTSS1}, and \textit{COL5A1\&2}; (\textbf{Group 2}) \textit{CD9} and \textit{CD74}; (\textbf{Group 3}): \textit{JUP}. The summary of the interactions among discovered biomarker-verse to cancer phenotypes shows disease progression and metastasis causality. We find potential pocket sites for each target marker with data from PDB\cite{berman2002protein} protein bank. Two pocket sites for targeting are found on large molecular \textit{MTSS1} and \textit{JUP}, while two binding sites are predicted in \textit{CD9} and one site in \textit{CD74}.}
    \label{fig:cd+}
\end{figure}
We translate the clinical findings toward potential therapeutic targets for cancer treatments. We use COSMIC\cite{tate2019cosmic} database to identify potential binding sites over the discovered biomarkers. \textbf{Figure}~\ref{fig:cd+} shows the analysis on three group of biomarkers reported in \textbf{Table}~\ref{tab:res_1} and ~\ref{tab:gene-pathways}. The interactions between biomarkers and cancer phenotypes have been summarized to demonstrate the causal relationship between disease progression and metastasis. By utilizing data from the protein data bank (PDB\cite{berman2002protein}), potential pocket sites for each target marker have been identified. Two pocket sites have been discovered on the large molecular \textit{MTSS1} and \textit{JUP}, while two binding sites have been predicted in \textit{CD9} and one site in \textit{CD74}. A comprehensive report using $10$ projects\cite{nguyen2022genomic,icgc2020pan,wu2022landscape,bolton2020cancer,rosen2020trk,samstein2019tumor,hyman2018her,miao2018genomic,robinson2017integrative,zehir2017mutational} from cBioPortal\cite{cerami2012cbio} on the structural information with mutational profiles of the complex \textit{COL5A1,2,3} and \textit{CD9-CD34-CD74} is given in \textbf{Figure}~\ref{fig:lollipop_1}; and that of CCT5, MTSS1 and NCAPD2 is given in \textbf{Figure}~\ref{fig:lollipop_2}.

In \textbf{Figure}~\ref{fig:lollipop_1}, \textit{CD9} has a common mutation at A106V, which involves two small sub-structures, while \textit{CD34} has a hot spot mutation at R314H/C, which is a large complex of antigen. In \textit{COL5A2} and \textit{COL5A3}, we found two regions with high mutation rates in the COLFI (Fibrillar collagen C-terminal domain). In contrast, \textit{COL5A1}'s hot spot mutation is at E267K near the laminin G-like domain. This complex is located in the extracellular matrix glycoproteins laminin, which binds to heparin and the cell surface receptor $\alpha$-dystroglycan.

In \textbf{Figure}~\ref{fig:lollipop_2}, we identified three potential sites with high mutation rates, such as the sub-unit CND1 of \textit{NCAPD2}, which tends to be mutated in both ends, with more mutations recorded at the R1241H/S/C end. Another potential target is the WH2 domain of \textit{MTSS1}, a 35 residue actin monomer-binding motif that plays a central role in many cell biological processes. We also found sites with high mutations recorded in drosophila immune deficiency (IDM) of \textit{MTSS1}, but the binding configurations were not provided in these databases\cite{berman2002protein}. We recommend studying these potential targets in the future.

\section{Discussion}\label{sec:discussion}
\subsection{Quant\textit{Ants} Machine is A Machine Learning Algorithm}
The input data of Quant\textit{Ants} machine is the two network domains, and the output is the pheromone trait maps associated with the G-Score of evaluated biomarker sets. Thus, the model outcome is data-driven; thus, Quant\textit{Ants} machine is a Machine Learning algorithm. The proposed model is generalized; however, the algorithm's scalability by several input genes should be addressed in further studies.

\subsection{Quant\textbf{Ants} Machine is A Quantum Algorithm}
The core of Quant\textit{Ants} Machine algorithm is the defined representations in \textbf{Definition}~\ref{def:dual_reps}. Of note, this representation is equivalent to unitary transformations in quantum computing. Specifically, we have the equivalent class of representations\cite{steinberg2012representation}
\begin{equation}
    \phi = \begin{bmatrix}
        \cos \bigg (\frac{2\pi m}{n} \bigg) & -\sin \bigg (\frac{2\pi m}{n} \bigg)\\
        \sin \bigg (\frac{2\pi m}{n} \bigg) & \cos \bigg (\frac{2\pi m}{n} \bigg)\\
    \end{bmatrix}
    \equiv \begin{bmatrix}
        \exp(\frac{2\pi i m}{n}) & 0 \\
        0 & \exp(-\frac{2\pi i m}{n}) \end{bmatrix},
\end{equation}
where $\phi$ is the conventional Ry-rotation of gate-based quantum computers\cite{nguyen2022bayesian,nguyen2022quantum}.

\section{Conclusion}\label{sec:conclusion}
To this end, I have introduced Quant\textit{Ants} machine - a quantum algorithm for biomarker discovery in \textbf{Section}~\ref{sec:method}. Two algorithms to run single and multi-ant systems are introduced in \textbf{Algorithm}~\ref{algo:quantants} and ~\ref{algo:multi-quantants}. The numerical results address the emergent \textit{RAS}-family target, which is accounted for many cancer types. Quant\textit{Ants} elucidates biomarkers that are rarely known by clinical literature, comprehensively reported in \textbf{Table}~\ref{tab:res_1} and ~\ref{tab:gene-pathways}. Finally, the study concluded by the translation of model outcome to potential therapeutics, which is summarized in \textbf{Figure}~\ref{fig:cd+}, \textbf{SuppFig}~\ref{fig:lollipop_1} and ~\ref{fig:lollipop_2}.

\bibliography{output}
\bibliographystyle{abbrv}

\begin{sidewaystable}[h]
\centering
\caption{\textbf{Gene Information and Pathways of Newly Discovered Biomarkers based on Comprehensive Literature Mining.} The clinical-known, associated pathways are summarized from\cite{griffith2017civic}.}
\label{tab:gene-pathways}
\begin{tabular}{|p{2cm}|p{2cm}|p{3cm}|p{14cm}|}
\hline
\textbf{Gene symbol} & \textbf{Gene name} & \textbf{Description} & \textbf{Pathways} \\ \midrule
\textit{CCT5} & T-complex protein one subunit epsilon & Molecular chaperone that is a chaperonin with TCP1 complex (CCT), also known as the TCP1 ring complex (TRiC) & Organelle biogenesis and maintenance, Prefoldin mediated transfer of substrate to CCT/TriC, Cooperation of Prefoldin and TriC/CCT in actin and tubulin folding, Formation of tubulin folding intermediates by CCT/TriC, Folding of actin by CCT/TriC, Chaperonin-mediated protein folding, Association of TriC/CCT with target proteins during biosynthesis, Protein folding, Metabolism of proteins, Assembly of the primary cilium, Cargo trafficking to the periciliary membrane, BBSome-mediated cargo-targeting to the cilium \\ \midrule
\textit{COL5A1} & Collagen alpha-1(V) chain & Minor connective tissue component of nearly ubiquitous distribution & Integrins in angiogenesis, Validated transcriptional targets of deltaNp63 isoforms, Beta1 integrin cell surface interactions, Syndecan-1-mediated signaling events, Extracellular matrix organization, Collagen formation, collagen biosynthesis, and modifying enzymes, Allograft rejection, miRNA targets in ECM and membrane receptors, Focal Adhesion, PI3K-Akt signaling pathway \\ \midrule
\textit{COL5A2} & Collagen alpha-2(V) chain & Minor connective tissue component of nearly ubiquitous distribution & Integrins in angiogenesis, Validated transcriptional targets of deltaNp63 isoforms, Beta1 integrin cell surface interactions, Syndecan-1-mediated signaling events, Extracellular matrix organization, Collagen formation, collagen biosynthesis, and modifying enzymes, Allograft rejection, miRNA targets in ECM and membrane receptors, Focal Adhesion, PI3K-Akt signaling pathway \\ \midrule
\textit{MTSS1} & Metastasis suppressor protein 1 & Could be related to cancer progression or tumor metastasis & Hedgehog signaling events mediated by Gli proteins \\ \midrule
\textit{NCAPD2} & Condensin complex subunit 1 & Regulatory subunit of the condensin complex used in the interphase chromatin's conversion into mitotic-like condense chromosomes & Aurora B signaling, Akap95 role in mitosis and chromosome dynamics, Cell Cycle, condensation of prometaphase chromosomes, mitotic prometaphase, M Phase \\ \midrule
\end{tabular}
\end{sidewaystable}

\begin{appendices}
\begin{figure}[h]
    \centering
    \includegraphics[width = 0.7\textwidth]{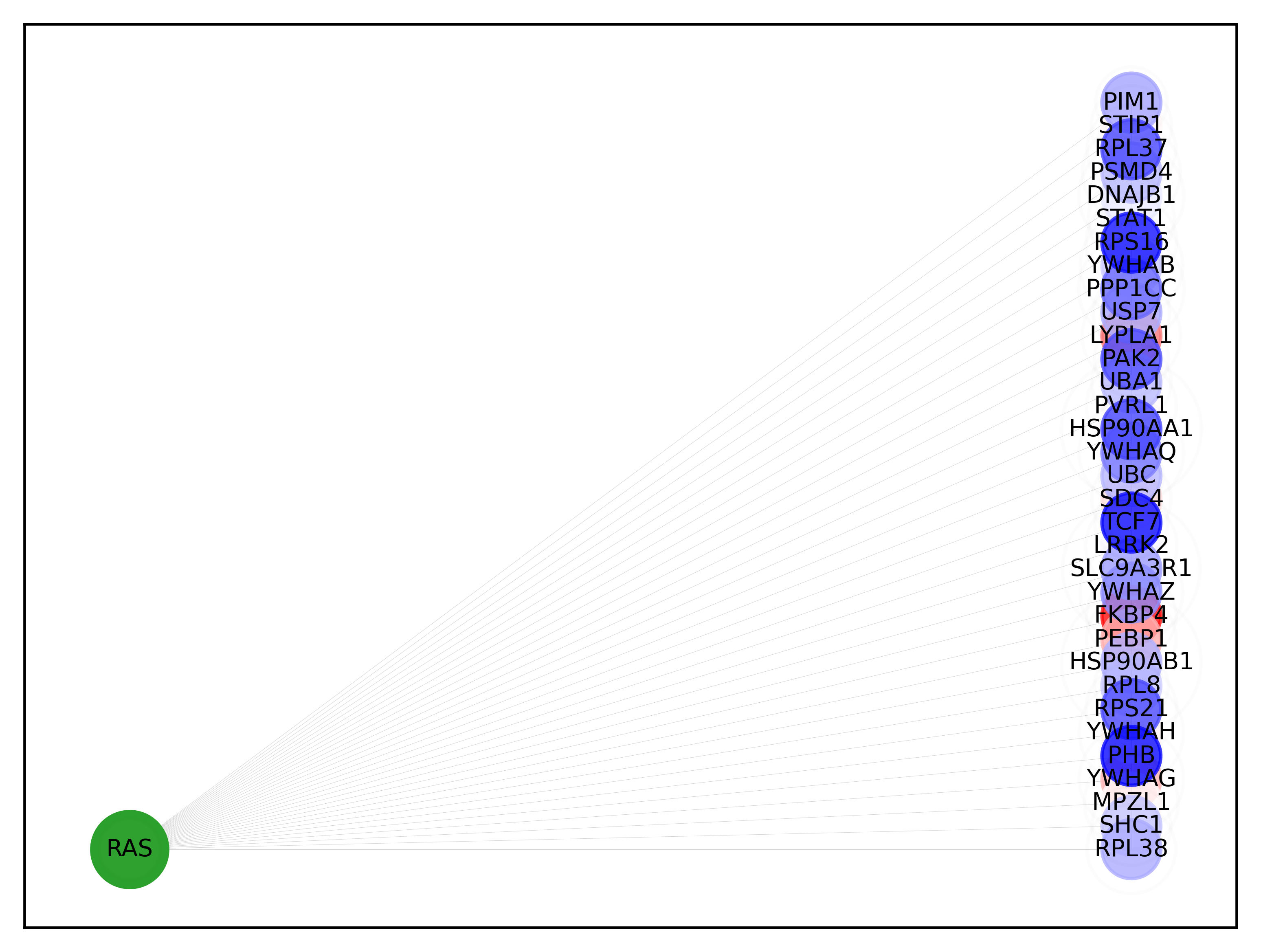}
    \includegraphics[width = 0.7\textwidth]{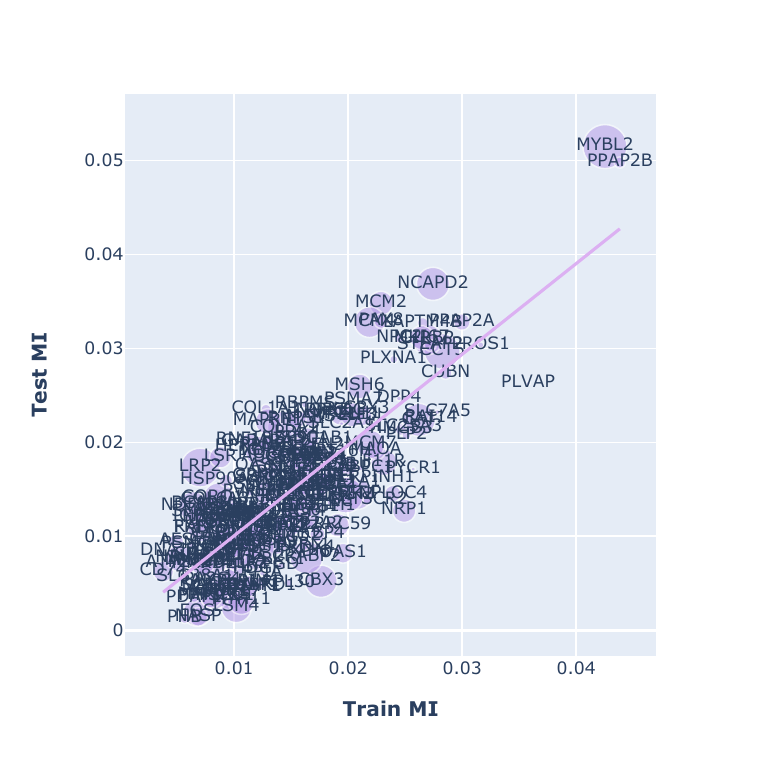}
    \caption{(Top):\textbf{The Interaction Graph generated by mRMR Criteria.} Highly relevant biomarkers are in blue, otherwise colored as red. (Bottom) \textbf{Train/Test MI to Targeted Signal RAS Activated Mutations Computed from TCGA-RNA Dataset}\cite{weinstein2013cancer}. We show a verse of top genes with test scores larger than global mean scores. Genes with higher pheromone intensities left by Quant\textbf{Ants} are highlighted in larger markers. The correlation $R$-squared between train and test MI is $0.64$, showing relatively good generalization of the discovered marker combination.}
    \label{fig:mrmr_graph}
\end{figure}
\begin{figure}[h]
    \centering
    \includegraphics[width = \textwidth]{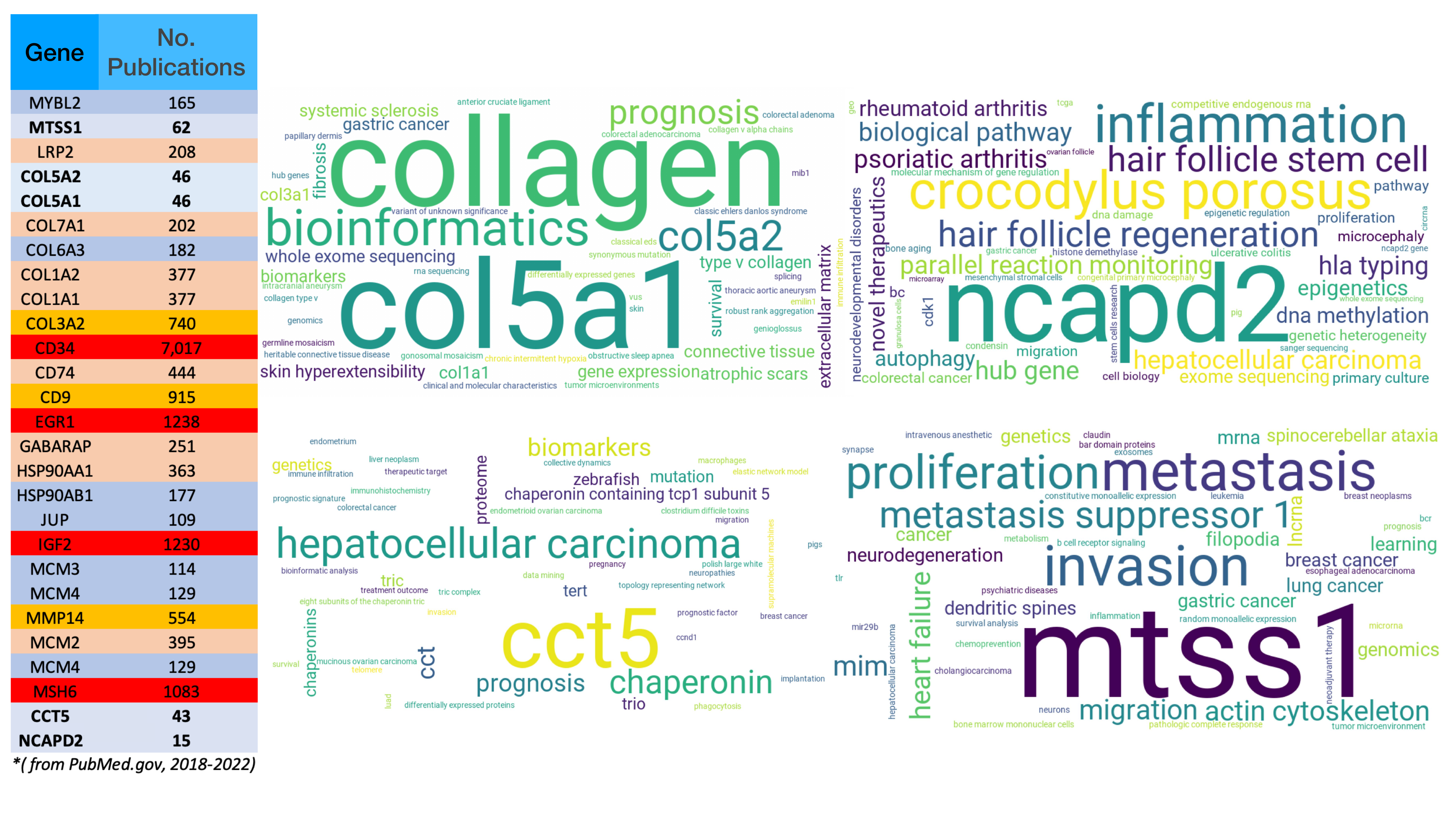}
    \caption{\textbf{Clinical Significance Summarized by Literature Mining using  \textit{PubMed.gov} Library, during The Period 2018-2022 (up to \textit{12/06/2022})} We mined $16,661$ publications regarding $27$ notable markers. In bold text, we highlight the biomarker discussed in less than $100$ papers. The word-cloud plots are generated by TopicTracker\cite{top_track}, associated with some biomarkers.}
    \label{fig:cli_sig}
\end{figure}
\begin{figure}[htb]
    \centering
    \includegraphics[width = \textwidth]{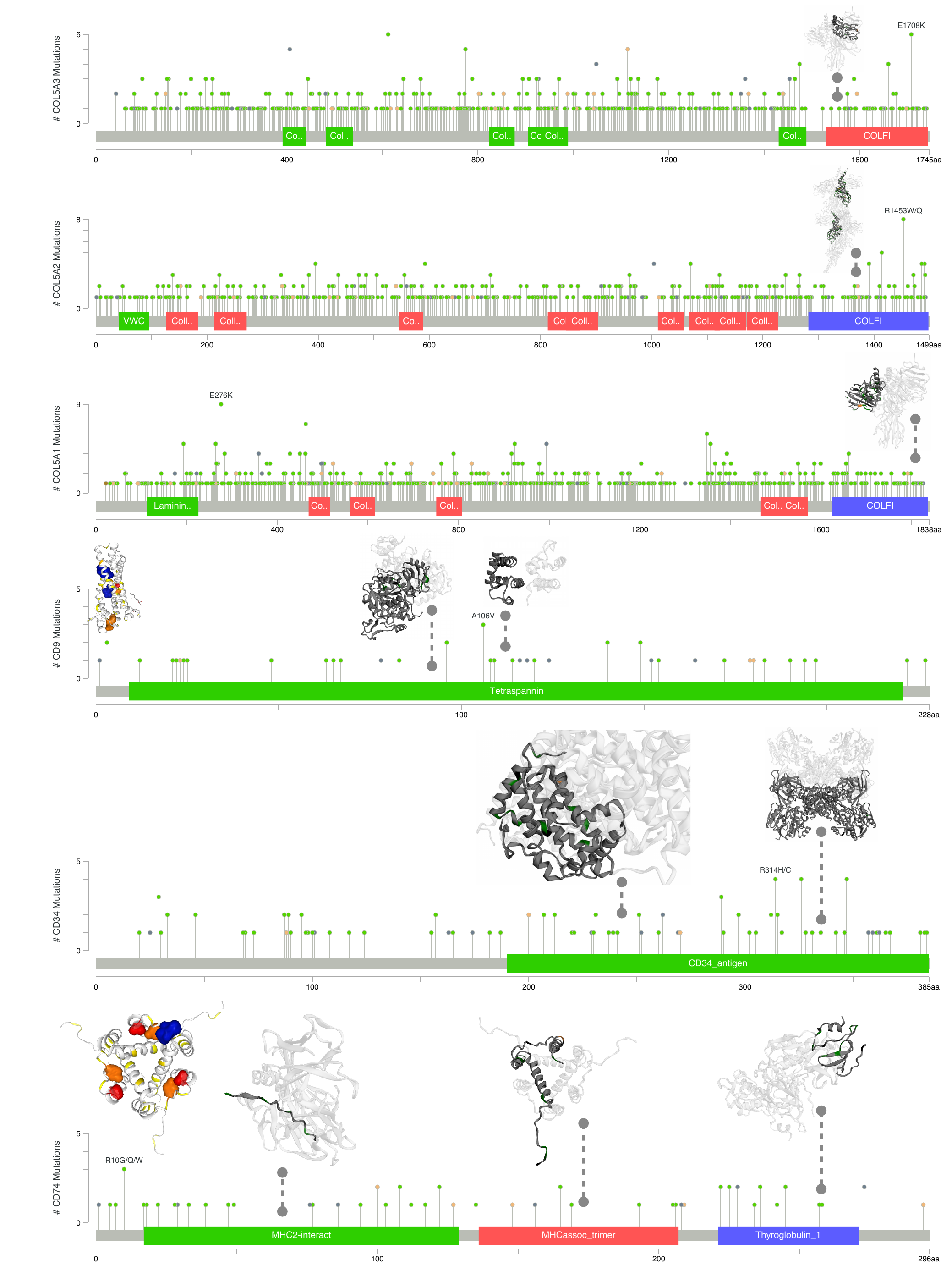}
    \caption{\textbf{Protein Structures of the Founded Markers with Mutational Profiles by Analyzing cBioPortal\cite{cerami2012cbio} Projects}. Part 1 - Small targets: \textit{CD9-34-74} and \textit{COL5A1-2-3}. There are notable sites with high mutation rates in which structures are not reported from the COSMIC database\cite{tate2019cosmic}.}
    \label{fig:lollipop_1}
\end{figure}

\begin{figure}[htb]
    \centering
    \includegraphics[width = \textwidth]{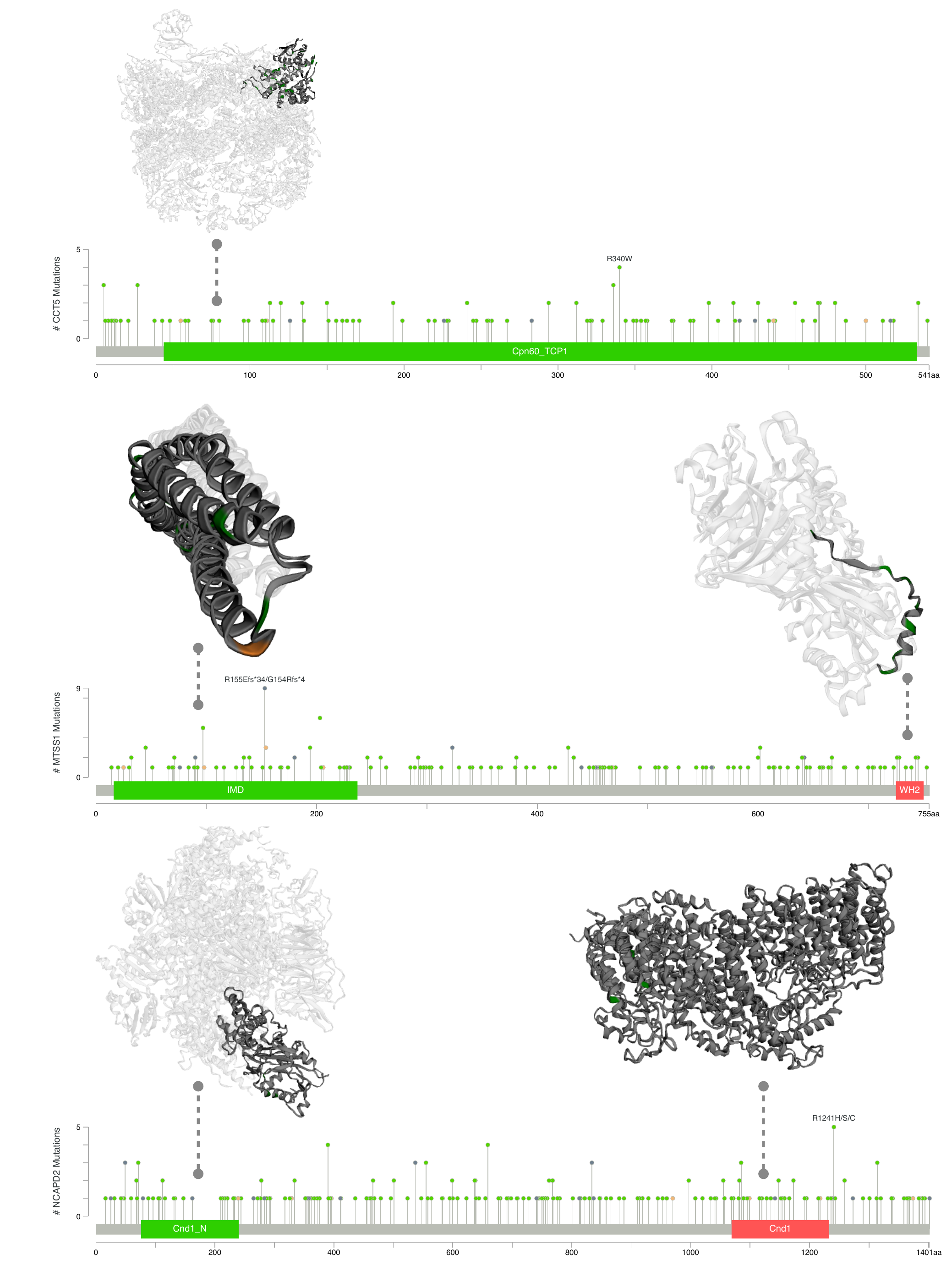}
    \caption{\textbf{Protein Structures of the Founded Markers with Mutational Profiles by Analyzing cBioPortal\cite{cerami2012cbio} Projects}. Part 2 - Large targets: \textit{CCT5}, \textit{MTSS1} and \textit{NCAPD2}. There are notable sites with high mutation rates in which structures are not reported from the COSMIC database\cite{tate2019cosmic}.}
    \label{fig:lollipop_2}
\end{figure}

\section{Discussion}\label{supp:discussion}

\textbf{Motivation:} Quant\textit{Ants} Machine is the quantum analogy of the well-known classical ant colony optimization (ACO). The ACO algorithm is a discrete optimization algorithm introduced in \cite{dorigo2006ant}, which has the natural beauty of nature: a group of ants walking on a graph $\mathcal{G}$ (\textbf{domain}) from their nest (site 1) to the food source (site 2) and then come back to the nest (site 3). During the process, they left \textbf{pheromone trait} on their paths. Suppose each ant carries an equal initial tank of liquid, decaying during the quest from (site 1) to (site 2) to (site 3). The intensity is computed by (liquid volume)/(total distance). Thus, the longer the distance, the less intensive the pheromone; equivalently, the shorter path will have a higher pheromone intensity. 

Ants can sense the pheromone trait to build stochastic models for the optimal traveling paths: 
\begin{itemize}
    \item The pheromone is evaporating due to the environment's evolution. Thus, a path with a more intensive pheromone will last longer, and more ants will follow these paths. This phenomenon is beneficial-natural bias; the system of ants is optimized in the sense of resources: more ants will have to travel less due to the optimal pheromone trait paths.
    \item When ants touch each other, there is hypothesized to have communication transferred from ant to ant. The biological facts are out of the scope of this work. Here, we observe that if the information can be shared among ant workers, it will be more optimized and beneficial, as they will share prior knowledge so that (1) more-common paths will be known and (2) less-common paths will be forgotten.
\end{itemize}

We introduce the quantum analog of ACO algorithm, named Quant\textit{Ants} Machine for biomarker discovery in oncology to achieve two major goals:
\begin{enumerate}
    \item Quant\textit{Ants} Machine overcomes the computational challenges of biomarker discovery for classical computing (\textbf{Section}~\ref{sec:introduction}) since computational acceleration can be achieved due to simultaneous sampling (\textbf{Algorithm}~\ref{algo:quantants} and ~\ref{algo:multi-quantants}).
    \item Quant\textit{Ants} Machine enables biomarker discovery over multiple modalities of cancer development by learning from dual-domain representations (\textbf{Equation}~\ref{equa:lambda_phi}).
\end{enumerate}

\textbf{Previous Submission:}
Quant\textit{Ants} Machine was submitted to Nature Cancer (NATCANCER-A08553) with the associated pre-print\cite{nguyen2022quantants}. We summarize the merit comments of reviewers in two major points. The development of quantum algorithms for identifying biomarker combinations will be compelling to researchers in this field. However, the conceptual and translational advance these findings represent for a broad cancer research audience is more limited. Thanks to the valuable reviews, I decided to re-organize the article. The summary of changes is as follows:
\begin{enumerate}
    \item The \textbf{Algorithm}~\ref{algo:quantants} and ~\ref{algo:multi-quantants} are moved to the main text to target mathematical and computational audiences.
    \item Clinical findings is comprehensively summarized in \textbf{Table}~\ref{tab:res_1} and ~\ref{tab:gene-pathways}.
    \item \textbf{Figure}~\ref{fig:lollipop_1} and ~\ref{fig:lollipop_2} is adapted and moved into \textbf{Appendix}.
    \item Minor adaptations in \textbf{Figure}~\ref{fig:maps}, ~\ref{fig:result_1} and ~\ref{fig:cd+} with the same analysis with the previous submission. Figure~\ref{fig:mrmr_graph} are added to provide preliminary details.
    \item The sgRNA design is excluded and will be introduced in a separate article.
    \item In general, we revise the article for a broader audience, ranging from computational and clinical practitioners.
\end{enumerate}

\section{Technical Reports}\label{supp:tech_rep}
\subsection{mRMR Graph Generation}\label{supp:mrmr}
We categorize pathways of interest in cancer research into two types: (1) direct or (2) indirect pathways. In the former case, the target variable $\bm{Y}$ straightforwardly presents the appearance of cancer types or sub-types (disease class). In contrast, the latter case concerns genetic-signaling pathways underlying disease-of-interest, such as TP53\cite{way2018machine} or RAS activation\cite{knijnenburg2018genomic}. We aim to identify a subset of genes $\mathcal{G} = \{G_i\}_{i\in I = \{1,\dots, d\}}$ which is considered $d$-biomarker for the targeted pathways $\bm{Y}$. The selection criteria for gene $G_i$ is based on mutual information between the gene's molecular feature $\bm{X}_i$ (DNA methylation, mRNA or RNA expression) and target $\bm{Y}$, given by\cite{nielsen2002quantum}
\begin{equation}\label{equa:I_XY}
    I(\bm{X}_i; \bm{Y}) = H(\bm{X}_i:\bm{Y}) = H(\bm{X}_i) - H(\bm{X}_i|\bm{Y})
\end{equation}
where 
\begin{equation}\label{equa:H_X}
     H(\bm{X}_i) = -\sum_x p_x \log p_x.
\end{equation}
is the Shannon entropy of a random variable $\bm{X}_i$. The calculation of the IT-quantity in Equation~\ref{equa:I_XY} is challenging in multivariable case $\mathbb{S} = \{\bm{X}_1, \dots, \bm{X}_k \}$ since the estimation of the joint distribution $p(\bm{X})$ is computationally expensive with current techniques. We can reduce the difficulty of the problem by assuming that $\bm{X}_i$ are independent variables, yielding $p(\bm{X}_1,\dots,\bm{X}_k) = \prod_{k}p(\bm{X}_k)$. However, this is a weak assumption for genomics features since we argue that the biological interactions among genes in set $\mathcal{G} \{G_i\}_{i\in I = \{1,\dots, k\}}$ straightforwardly induce statistical dependency among their feature set $\bm{X} = \{\bm{X}_1,\dots,\bm{X}_k\}$.

An efficient method to reduce the computational complexity of the given problem is using forward stage-wise search with \textit{relevancy} and \textit{redundancy} measurements. Specifically, the subset of features $\mathbb{S}$ is defined by greedy approaches, in which only a single gene $G_i$ is added to the subset $\mathbb{S}$ in each iteration. The Minimum Redundancy-Maximum Relevancy-based\cite{peng2005feature} (mRMR-based). Specifically, we aim to simultaneously maximize the relevancy between a new feature $\bm{X}_i$ with target $\bm{Y}$
\begin{equation}\label{equa:rel}
    \text{REL}(\bm{X}_i,\bm{Y}) := I(\bm{X}_i,\bm{Y})
\end{equation}
and 
minimizing the redundancy of the chosen set $\mathbb{S}$\cite{nguyen2014effective}
\begin{equation}\label{equa:red}
    \text{RED}(\bm{X}_i|\mathbb{S}) := \sum_{\bm{X}_j \in \mathcal{S}} I(\bm{X}_i,\bm{X}_j)
\end{equation}
The optimization problem is given as
\begin{equation}
    \max_{\bm{X}_i \in \mathcal{X} \setminus \mathbb{S}} = \{\text{REL}(\bm{X}_i)-\text{RED}(\bm{X}_i|\mathbb{S}) \}.
\end{equation}
This problem is equivalent to minimizing the loss value:
\begin{equation}
    \mathcal{L} = I(\bm{X}_i, \bm{Y}) - \lambda \sum_{\bm{X}_j \in \mathbb{S}} I(\bm{X}_i, \bm{X}_j),
\end{equation}
where $\lambda = 1/|\mathbb{S}|$ is considered weights that scales the pairwise-mutual information between $\bm{X}_i$ and $\bm{X}_j$. We show the interaction network of the RAS family (including NRAS, HRAS, and KRAS) computed by mRMR in \textbf{SuppFig}~\ref{fig:mrmr_graph}.

\subsection{Literature Mining}
We summarize the most used keywords regarding the biomarker found in \textbf{Table}~\ref{tab:res_1} and ~\ref{tab:gene-pathways} as follows
\begin{itemize}
    \item \textbf{CCT5}: \textbf{zebrafish}, clostridium difficile toxins, glycosyltransferase toxins, \textbf{hepatocellular carcinoma}, \textbf{tert}, dendrite morphogenesis, genetic chaperonopathies, endometrioid/mucinous \textbf{ovarian carcinoma}, \textbf{immunohistochemistry}, telomere, \textbf{breast cancer}, endometrium, \textbf{migration}, \textbf{therapeutic target}, \textbf{invasion}, \textbf{liver neoplasm}, \textbf{treatment outcome}, \textbf{colorectal cancer}, \textbf{immune infiltration}, \textbf{p53} (\textbf{TP53}), enzyme activity, \textbf{sperm proteome}, \textbf{sperm}, \textbf{fertility}, cancer stem cells.
    
    \item \textbf{COL5A1} and \textbf{COL5A2}: systemic sclerosis, \textbf{gastric cancer}, \textbf{COL3A1}, extracellular matrix, \textbf{COL1A1}, connective tissue, skin hyperextensibility, anterior cruciate ligament, \textbf{germline/gonosomal mosaicism}, \textbf{colorectal adenocarcinoma}, \textbf{colorectal adenoma}, cell adhesion, \textbf{inflammatory disease}, predictive biomarkers, \textbf{COL6A3}, \textbf{COL11A1}, \textbf{colorectal cancer}.
    
    \item \textbf{MTSS1}: \textbf{invasion}, \textbf{metastasis}, \textbf{proliferation}, metastasis suppressor 1, \textbf{migration}, \textbf{lung} cancer, \textbf{gastric} cancer, \textbf{ovarian} cancer, \textbf{breast} cancer, \textbf{brain} cancer, \textbf{hepatocellular carcinoma}, \textbf{breast} neoplasms, \textbf{neoadjuvant therapy}, \textbf{collagen I}, \textbf{immune infiltration}, genetic differentiation, \textbf{colorectal} cancer, \textbf{anxiety}, \textbf{weight} \textbf{loss}, \textbf{obesity}, \textbf{oncogene}, \textbf{acute myeloid leukemia}, cell growth, prostate cancer, DNA methylation, receptor protein tyrosine phosphatase, tumor suppressor.
    
    \item \textbf{NCAPD2}: inflammation, Crocodylus porosus, hair, follicle regeneration, hair follicle stem cell, \textbf{hepatocellular}, carcinoma, DNA methylation, novel therapeutics, bc, \textbf{CDK1}, \textbf{migration}, \textbf{proliferation}, \textbf{colorectal cancer}, DNA damage, \textbf{gastric cancer}.
\end{itemize}

We find the potential cancer phenotypes and some symptoms discovered within the literature mining. Moreover, we analyze additional drivers with potential therapeutic applications, including:
\begin{enumerate}
    \item \textbf{CD74} - HLA class II histocompatibility antigen gamma chain ($444$ papers): macrophage migration inhibitory factor, inflammation, tumor microenvironment, \textbf{immunotherapy}, \textbf{lung} adenocarcinoma, \textbf{breast} cancer, \textbf{immunohistochemistry}, \textbf{migration} \textbf{inhibitory} \textbf{factor}, \textbf{innate} \textbf{immunity}, proliferation, \textbf{pancreatic} cancer, \textbf{hepatocellular} carcinoma, \textbf{metastasis}, \textbf{MAPK}, \textbf{fish}, \textbf{TP53}, \textbf{KRAS}, \textbf{metastatic} \textbf{melanoma}, \textbf{cervical} cancer, \textbf{AKT}, \textbf{epigenetic}, \textbf{immune} \textbf{resistance}, \textbf{cleaner} \textbf{fish}, \textbf{T-cell}, \textbf{metastatic} \textbf{melanoma}, \textbf{immunotherapy} \textbf{resistance}, driver mutation, \textbf{T-cell activation}, \textbf{immunesuppresion}, \textbf{adaptive} \textbf{immune} \textbf{response}, \textbf{EGFR} \textbf{mutation}, \textbf{immune} \textbf{infiltration}.

    \item \textbf{JUP} - Common junctional plaque protein ($109$ papers): inflammation, \textbf{gastric} cancer, \textbf{prostate} cancer, wnt family member 10a, gamma catenin, \textbf{KPP}, \textbf{GJB6}, \textbf{GJB2}, \textbf{EGFR}, \textbf{MTBPS2}, \textbf{PPK}, \textbf{TP53} apoptosis effector related to pmp22, oncogenes, apoptosis, \textbf{UGDH}, \textbf{PDGFRB}, \textbf{HL1}, \textbf{CRISPR}, \textbf{bladder} cancer.
\end{enumerate}

\subsection{Identification of Therapeutic Targets via Biomarker Structural Analysis}
Analyses on $10$ databases\cite{nguyen2022genomic,icgc2020pan,wu2022landscape,bolton2020cancer,rosen2020trk,samstein2019tumor,hyman2018her,miao2018genomic,robinson2017integrative,zehir2017mutational} from cBioPortal\cite{cerami2012cbio} with $76,639$ samples, we obtain the mutational profiles of potential targets discussed in \textbf{Section}~\ref{sec:clinical_significance}, summarized in the lollipop plots \textbf{SuppFig}~\ref{fig:lollipop_1} and ~\ref{fig:lollipop_2}.

\end{appendices}

\end{document}